\documentclass[conference]{IEEEtran}
\usepackage{cite}
\usepackage{amsmath,amssymb,amsfonts}
\usepackage{algorithmic}
\usepackage{graphicx}
\usepackage{textcomp}
\usepackage{xcolor}
\usepackage{refstyle}
\usepackage[numbers]{natbib}
\def\BibTeX{{\rm B\kern-.05em{\sc i\kern-.025em b}\kern-.08em
    T\kern-.1667em\lower.7ex\hbox{E}\kern-.125emX}}
\usepackage[bookmarks=false,colorlinks=true,linkcolor=blue]{hyperref}

\IEEEoverridecommandlockouts
\IEEEpubid{\makebox[\columnwidth]{978-1-7281-6251-5/20/\$31.00~\copyright2020 IEEE \hfill} \hspace{\columnsep}\makebox[\columnwidth]{ }}

\begin{document}
\bstctlcite{IEEEexample:BSTcontrol}

\title{Deep Learning for Surface Wave Identification\\
  in Distributed Acoustic Sensing Data
  \thanks{This work is supported by the Laboratory Directed Research and Development (LDRD) Program of Lawrence Berkeley National Laboratory under U.S. Department of Energy Contract No.~DE-AC02-05CH11231. 
It used resources of the National Energy Research Scientific Computing Center and Energy Science network (ESnet), both are funded under the above contract.
The U.S. Government retains, and the publisher, by accepting the article for publication, acknowledges, that the U.S. Government retains a non-exclusive, paid-up, irrevocable, world-wide license to publish or reproduce the published form of this manuscript, or allow others to do so, for U.S. Government purposes.}}

\author{
    \IEEEauthorblockN{
        Vincent Dumont\IEEEauthorrefmark{1},
        Ver\'{o}nica Rodr\'{i}guez Tribaldos\IEEEauthorrefmark{1},
        Jonathan Ajo-Franklin\IEEEauthorrefmark{2}\IEEEauthorrefmark{1},
        and
        Kesheng Wu\IEEEauthorrefmark{1}
    }
    \IEEEauthorblockA{
        \IEEEauthorrefmark{1}
            \textit{
                Lawrence Berkeley National Laboratory,
                Berkeley, CA, USA
            }
        \IEEEauthorrefmark{2}
            \textit{
                Rice University,
                Houston, TX, USA
            }
    }
}

\maketitle
\IEEEpubidadjcol

\begin{abstract}
Moving loads such as cars and trains are very useful sources of seismic waves, which can be analyzed to retrieve information on the seismic velocity of subsurface materials using the techniques of ambient noise seismology. This information is valuable for a variety of applications such as geotechnical characterization of the near-surface, seismic hazard evaluation, and groundwater monitoring. However, for such processes to converge quickly, data segments with appropriate noise energy should be selected. Distributed Acoustic Sensing (DAS) is a novel sensing technique that enables acquisition of these data at very high spatial and temporal resolution for tens of kilometers. One major challenge when utilizing the DAS technology is the large volume of data that is produced, thereby presenting a significant Big Data challenge to find regions of useful energy. In this work, we present a highly scalable and efficient approach to process real, complex DAS data by integrating physics knowledge acquired during a data exploration phase followed by deep supervised learning to identify ``useful'' coherent surface waves generated by anthropogenic activity, a class of seismic waves that is abundant on these recordings and is useful for geophysical imaging. Data exploration and training were done on 130~Gigabytes (GB) of DAS measurements. Using parallel computing, we were able to do inference on an additional 170~GB of data (or the equivalent of 10 days' worth of recordings) in less than 30 minutes. Our method provides interpretable patterns describing the interaction of ground-based human activities with the buried sensors.
\end{abstract}

\begin{IEEEkeywords}
Big Data, Deep Learning, Distributed Acoustic Sensing, Distributed Training, Sensor Network Technology
\end{IEEEkeywords}


\section{Introduction}

\begin{figure}
\includegraphics[width=\columnwidth]{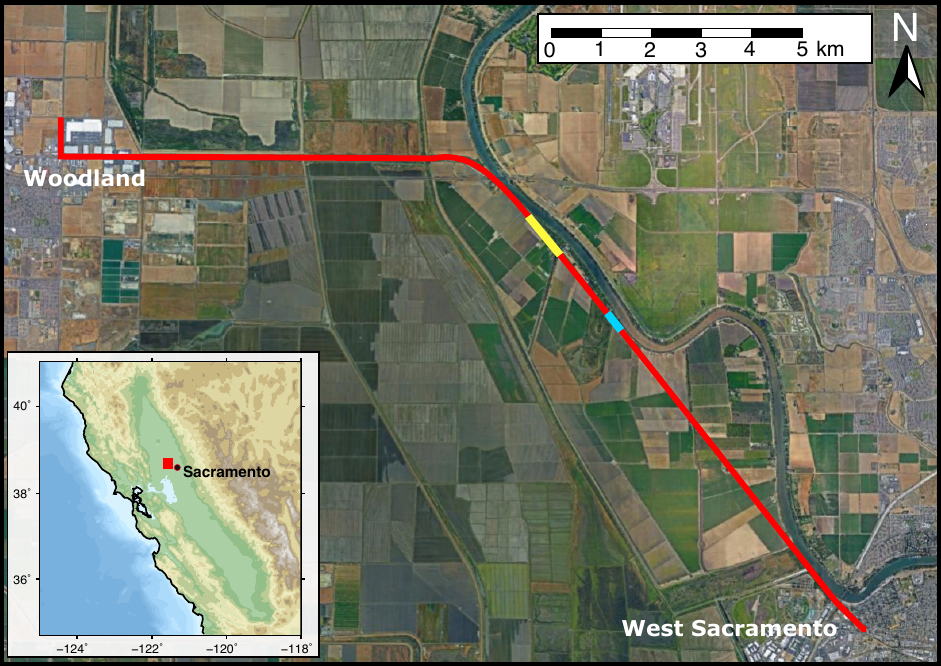}
\caption{Location of the optical fiber generating the data used in this work. The fiber, shown in red, is located North-West of Sacramento, California. The section of channels used as subsets A and B are shown in yellow. In blue shows the section for subset C.}
\label{fig:fiber_map}
\end{figure}


The increase use of sensor network technology to acquire data has turned many scientific research activities into Big Data applications~\citep{2017bowen,2018afach}. Deep learning, combined with distributed parallel computing, has proven to be a powerful tool to significantly speed up the analysis of these large scientific data sets. However, one challenge in applying these learning methods is the lack of labeled data for training. In this work, we utilize domain experts to devise a data exploration process to turn physics insight into usable labels for a complex set of scientific data from geoscience acquired with a new technology known as Distributed Acoustic Sensing (DAS).
Through this collaboration, we develop a set of efficient tools that can process hundreds of gigabyes of data in minutes. These tools utilize high-performance libraries such as \textsc{mpi4py} \cite{mpi4py} and \textsc{ArrayUDF} \cite{2017bin} to complete critical operations in supporting interactive operations on large amounts of data. These tools not only allow us to process data quickly, but also produce interpretable patterns relating signals to human activities.


Distributed Acoustic Sensing (DAS) is an emerging technology that effectively transforms conventional fiber-optic cables into massive arrays of single component seismometers that enable the acquisition of dense, high-resolution data sets across 10s of kilometers.
DAS uses a variant of phase-sensitive time-domain reflectometry ($\Phi$-OTDR) to make spatially distributed measurements of strain rate along fiber-optic cables \citep{Hartog2017}. DAS systems send consecutive, coherent laser pulses down a fiber-optic cable buried in the ground. Due to naturally-occurring impurities within the core of the fiber, part of the light sent down the cable is backscattered and recorded back at the DAS unit, where the optical phase of the light is measured.
Seismic waves impacting the cable induce compression/extension, causing a change in the optical phase of the backscattered light.
An interferometric system in the DAS unit measures the optical phase changes of consecutive backscattered light profiles, which are proportional to changes in longitudinal strain along the fiber.
Current DAS systems enable acquisition of strain rate at spatial samplings as small as 25 cm at rates in the tens of kHz range.



DAS can be deployed on so-called dark fiber networks, existing
telecommunication fiber-optic cables not currently used for data
transfer~\citep{Ajo-Franklin2019}. Recent studies have shown that the
analysis of ambient seismic noise generated from traffic infrastructure
such as cars, trains, etc.~recorded on these networks provides an
attractive approach for regional, high-resolution near-surface
characterization and continuous monitoring of subsurface processes for
long periods of time \citep{Ajo-Franklin2019,Tribaldos2019}. In
particular, surface waves (i.e. seismic waves that travel along the
surface of the Earth) recorded within this infrastructure noise can be
analyzed to retrieve seismic velocity structure of the near surface,
which provides valuable information about soil properties. For these
investigations it is crucial to identify coherent events from noise records.

Current approaches used by the geophysical community use simple metrics such as raw data amplitude to select or variably weight the ambient seismic noise recordings that contain high-amplitude surface waves, so that they can be subject to further processing. This selection process is time consuming, requiring a significant amount of user input, and discards a large amount of continuously recorded data that could contain usable information. This is particularly challenging in regional, long-term, multi-month DAS studies, in which data volumes can easily amount to several terabytes a day. These limitations currently make it challenging to efficiently handle and process DAS ambient seismic noise recordings. It can take up to several months to analyze the data sets and, in many cases, only part of the data is used. It is thus evident that innovative approaches are needed to manage these large data sets and provide a means to take advantage of the unprecedented spatial and temporal resolution provided by this new technology.

A limited number of recent studies have explored leveraging Machine Learning (ML) approaches for analysis and processing of DAS ambient seismic noise datasets. However, they differ from our approach in the type of signals that they target. In the majority of these studies, the main objective is event detection and classification, and the feature of interest is the complete seismic signal generated by a vehicle or person travelling past the array, which can be differentiated from other signals of interest such as earthquakes. For example, \citet{liu2020} develop a combined ML approach based on feature extraction for vehicle detection, classification and speed estimation. \citet{Jakkampudi2020} develop a convolutional neural network to automatically detect footstep signals in ambient seismic recordings from urban DAS arrays. In a similar approach, \citet{Huot2018} use a convolutional neural network to automatically detect car-generated seismic signals with the objective of removing them from the seismic recordings. In the majority of cases, these signals are complex and composed of a combination of useful seismic energy with other effects such as static deformation due to the vehicle load, optical effects originated by the recording instrument, etc. In contrast, our study focuses in the identification and quantification of the useful seismic waves embedded within the seismic signal generated by the vehicles, which is a much subtler feature that requires careful examination and extraction.

In this study, we apply ML approaches to optimize the analysis of DAS ambient noise data sets. The two-dimensional data are first explored in time and frequency domains, and as single images through unsupervised learning. We then search for and define appropriate metrics to generate a labeled training data set. Using distributed parallel computing, we develop a classifier capable of automatically identifying surface wave energy from raw ambient noise DAS recordings, and assign a probability of surface wave occurrence in each data file.

\begin{figure}[t]
\includegraphics[width=\columnwidth]{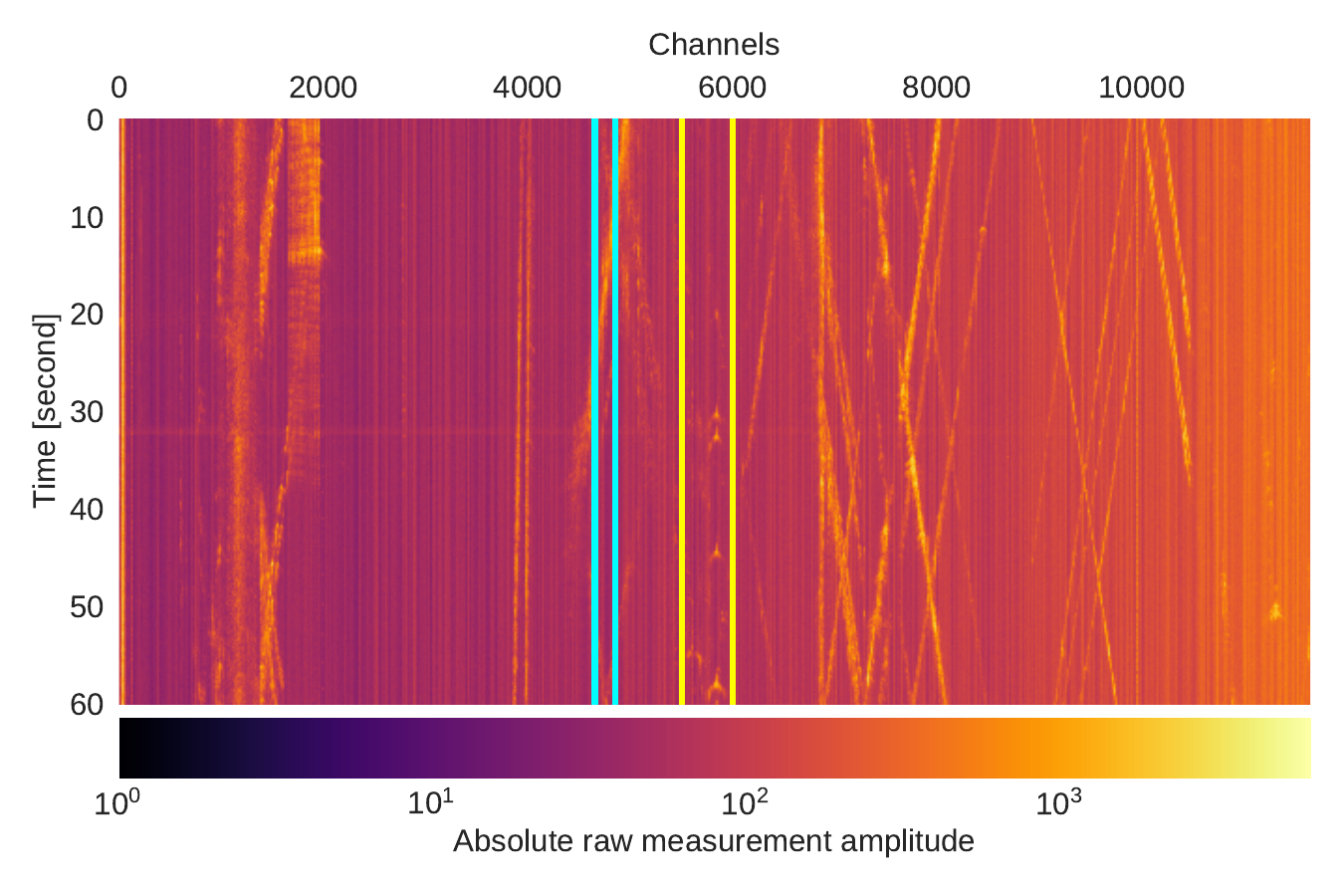}
\caption{One minute of DAS ambient seismic noise recorded over the entire optical fiber path (11,648 consecutive channels). Signal amplitude, which is proportional to strain-rate, is plotted in logarithmic scale. High-amplitude regions reflect areas with significant traffic seismic noise, rich on surface waves. Sections within yellow and blue vertical stripes show the fiber channel ranges from which data sets A, B and C have been extracted.}
\label{abs_data}
\end{figure}

Here, this method is applied to DAS data recorded as part of the Fiber-optic Sacramento Seismic Array (FOSSA) Dark Fiber experiment described in \citet{Ajo-Franklin2019} and shown in Figure~\ref{fig:fiber_map}. The fiber-optic cable is deployed in the subsurface, either directly buried in the ground or inside a conduit. It is roughly co-linear to a railway track surrounded by roads, both strong sources of anthropogenic seismic noise. The fiber is approximately 23~km long, and spatial sampling was 2~m, resulting in $\sim11,500$ channels (Fig. \ref{abs_data}). The experiment run between July 28th, 2017 and March 4th, 2018, and data was continuously acquired at a sampling frequency of 500~Hz, generating $\sim300$~TB of data. Data sets were recorded in the form of 1 minute-long files for the entire array. 

The key contributions of this work are:
\begin{itemize}
\item A bootstrap process to define a learning and inferencing workflow to effectively extract meaningful patterns from the large amount of DAS data.
\item A highly-efficient multi-GPU deep learning algorithm for classifying recorded signals in DAS data.
\item A scalable method to map the probability of coherent surface wave signals over large sets of DAS data within minutes.
\item An easy-to-use deep learning and data exploration software designed for DAS data and fully compatible on large scale computing environment.
\end{itemize}

\section{DAS data sets}

\begin{figure*}[t]
\includegraphics[width=\textwidth]{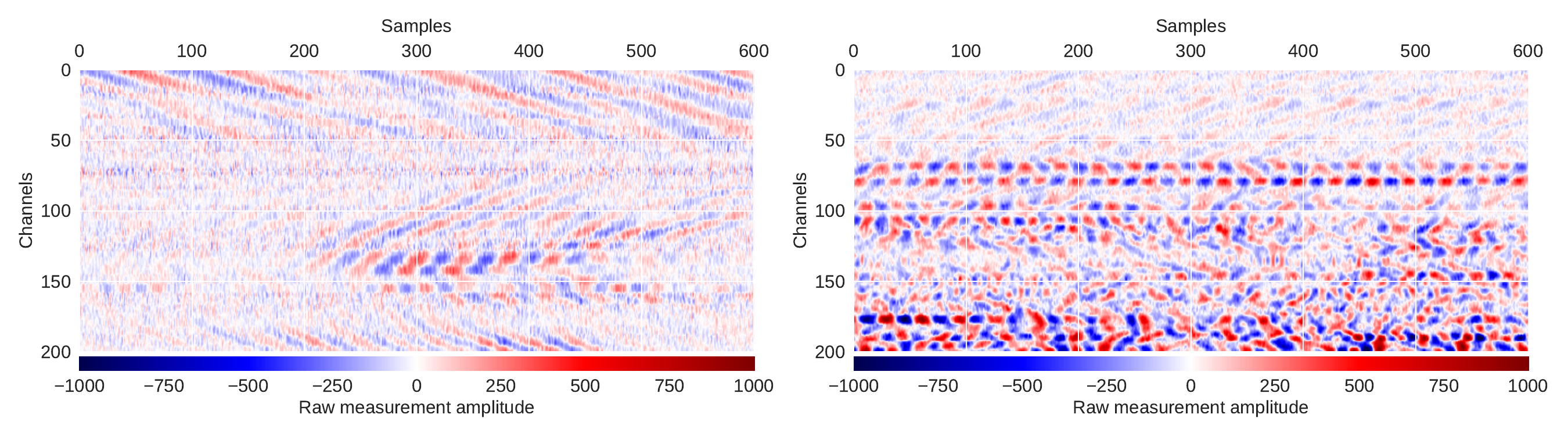}
\caption{Two examples of raw strain-rate measurements. The region on the left was taken from one of the files in data set A where train crossing occurs. The surface waves produced by the train can be observed at the top of the figure while an additional, shorter, surface wave signal can be seen in the bottom half of the region. The figure on the right originates from a 1-minute file from data set C. Shallow surface wave signals can be found in the top part of the figure. However, interferences between multiple surface wave signals appears to dominate the region.}
\label{raw_data}
\end{figure*}

DAS data sets consist of two-dimensional arrays, with a dimension of time (samples) and a dimension of space (measurement channels). Essentially, each channel corresponds to a time series at a specific location along the fiber. Because of the high density of measurements in both dimensions, DAS data sets can be treated as images for machine learning purposes. 
In this work, we analyse three different subsets of the same DAS data set (A, B and C), corresponding to different sections of the cable and/or different time periods.

{\bf Data set A} corresponds to 43 files containing 30-minute time series (i.e. 900k samples) from 500 consecutive measurement channels (5500 to 6000) recorded during the passing of trains near the fiber line between July 30\textsuperscript{th} and October 29\textsuperscript{th}, 2017. Trains are notable for producing coherent surface waves which can be used to characterize the sub-surface. This data set will be used for data exploration and generating labeled training images.

{\bf Data set B} corresponds to 103 files containing 30-minute time series data from 500 channels (5500 to 6000 included) recorded between August 3\textsuperscript{rd} and August 7\textsuperscript{th}, 2017, when no train was present. This data set can be used to explore the optical fiber background noise (that is, where no surface wave signals are found) and regions where seismic waves from other sources, such as vehicles, interfere to produce visually complex signal patterns. Like for data set A, this subset will be used during the data exploration phase and to generate the training images.

{\bf Data set C} corresponds to 14,400 files containing 1-minute time series data (i.e. 30k samples) from 200 consecutive channels (4650 to 4850 included) recorded over a period of 10 days, from January 1\textsuperscript{st} to 13\textsuperscript{th}, 2018. This data set will be used to do probability mapping using machine learning and identify useful energy signals in the data.

Each of the 30-minute files from data sets A and B contains a total of 450M strain measurements and each 1-minute data file from data set C contains 6M measurements. The amount of individual strain measurements recorded throughout the entire DAS data sample used in this work sums to over 152 billion measurements ($\sim$300\,GB), thereby making this research project a real Big Data challenge.

\section{Exploratory data analysis}

\subsection{Coherency of energy signals}

The main objective of this work is to identify the image pattern of surface waves as they travel along the fiber-optic cable. Fig. \ref{raw_data}, shows examples of raw data subsets from each of the fiber sections used in this study. In this time-space domain, ideal surface wave signals are visually recognized as "V-shaped" or linear features, which physically correspond to the waves stretching (positive amplitudes) and compressing (negative amplitudes) the fiber-optic cable. In many cases, however, these clean signals are not evident, and interfering surface waves and/or other processes impacting the cable can result in very complex signals that are not useful for geophysical analysis. Currently, there are no well-developed tools to efficiently identify useful surface wave energy in DAS data. Due to the complexity of the signals and the large volume of the data sets, a common practice is to select specific time periods or regions of the data in which these surface waves are expected to occur, such as a time at which a train is approaching the location of the cable, as the train is known to be a great source of seismic surface waves that will travel through the ground and impact the cable. In this work, we build a deep learning classifier designed to identify these usable surface waves and assign a probability of surface wave energy occurrence to any DAS recording. Labeled data sets are therefore needed to train the classifier, but the complexity of the data makes the preparation of these training data sets very challenging. During this exploration phase, we aim to identify the main types of signals that can be found in the data, and to explore their properties both in time and frequency domain. As a result, data metrics are selected and will be used to construct our training data set.

\subsection{Signal identification based on data distribution and signal amplitude}

\begin{figure}[ht]
\includegraphics[width=\columnwidth]{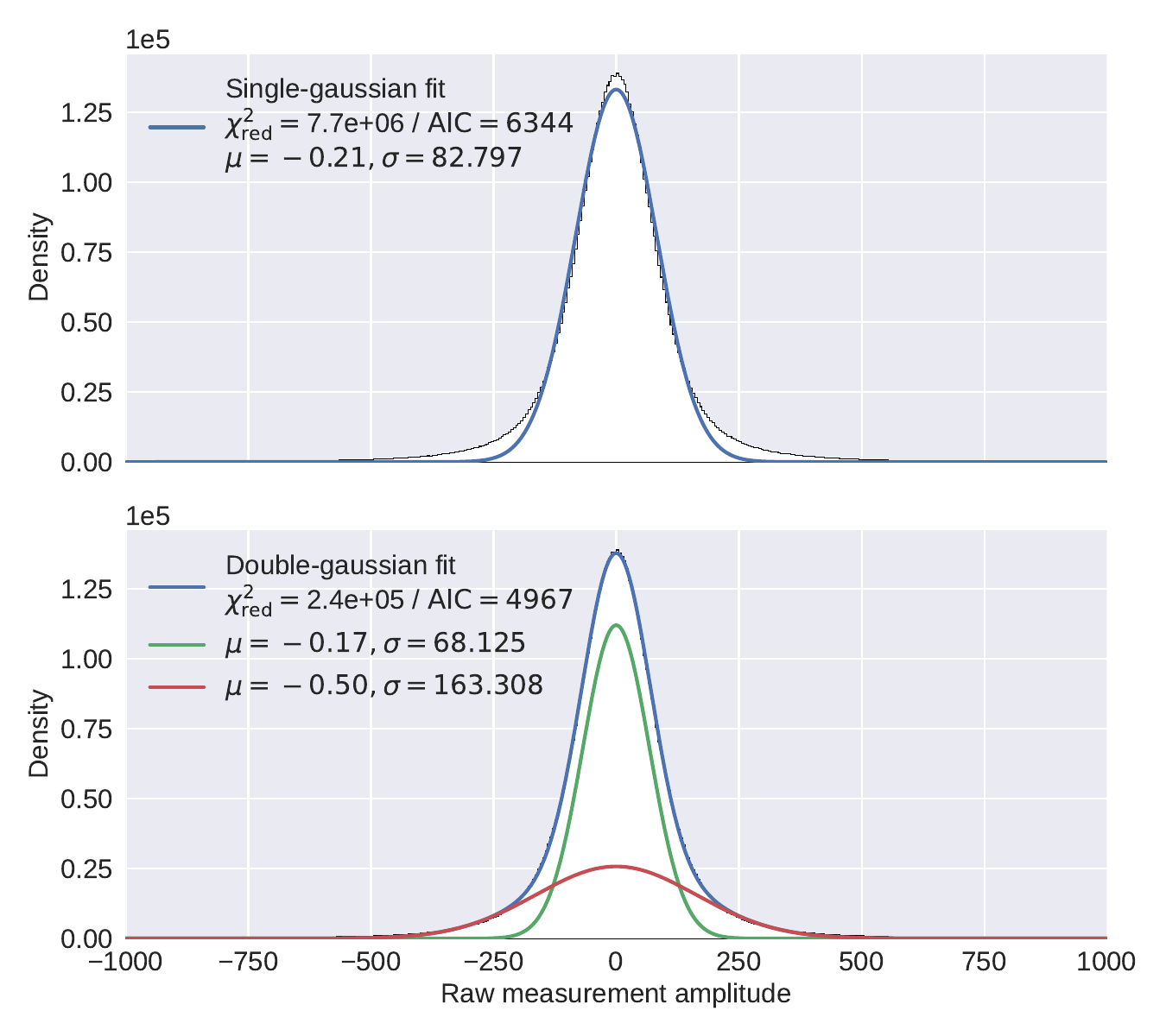}
\caption{Distribution of strain-rate measurements from a 1-minute, 200-channel DAS data set. A double-Gaussian fit to the distribution (bottom panel) provides a better approximation than using single-Gaussian fit (top panel). Multi-Gaussian fits can be used to identify different types of signals in the data. Here, the narrow component is representative of white noise signals while the broader component reflects the presence of higher energy signals.}
\label{distribution}
\end{figure}

One of the most evident data characteristics to explore is the amplitude of the recorded signals. The distribution of a 1-minute, 200-channel data region is shown in Fig. \ref{distribution}. Two fitting approaches of the distribution are explored, using single and double-Gaussian fit, respectively. The goodness of fit for both models can be estimated by calculating the reduced chi-square using the unweighted residuals as follows:
\begin{equation*}
    \chi^2_\mathrm{red}=\frac{1}{\mathrm{df}}\sum_i{(y_i-f(x_i))^2}
\end{equation*}

\noindent where $\mathrm{df}$ is the total number of degrees of freedom, equal to the sample size minus the number of fitted parameters (e.g. a single Gaussian model has 3 parameters: amplitude, mean, and sigma). We found that a double-Gaussian fit provides the best approximation. Each individual Gaussian curve is the representation of a particular signal pattern found in the data. The narrow Gaussian component, with a high probability density function (PDF) is characteristic of typical white noise, which has relatively low amplitude. On the other hand, the broader Gaussian component reflects areas where surface wave signals are found, with signal amplitude fluctuating from highly negative to highly positive values.

While the distribution of the amplitude of the raw measurements provides important insight into the types of signals available in the target region, it is not sufficient to build reliable selection criteria to distinguish features. For instance, regions with low-amplitude surface wave signals still carry useful energy because of their coherency. Nevertheless, low-amplitude signals will be very difficult to distinguish from purely white noise data as they are covering a similar range in the distribution. Moreover, looking back at Fig. \ref{raw_data}, one can see how surface waves quickly decrease in amplitude as they travel along the array, thereby making it even more challenging to clearly identify them solely based on data distribution. 

\subsection{Addition of frequency content of ambient noise signals}

\begin{figure}[h]
\includegraphics[width=\columnwidth]{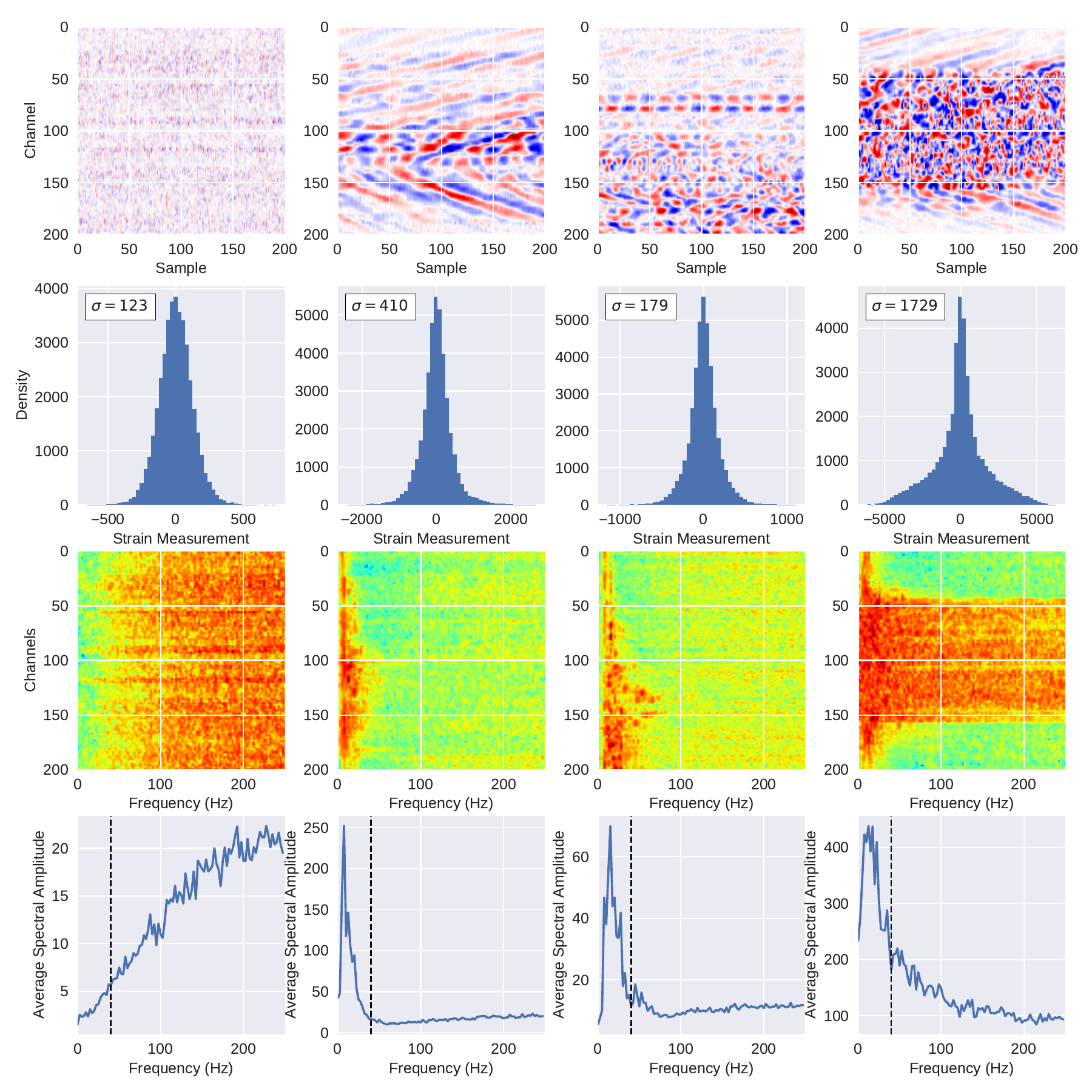}
\caption{Examples of signals found in DAS data. From left to right: (1) purely white noise region, (2) coherent surface wave signals, (3) non-coherent waves, and (4) saturated signal. The first row shows an example of raw data for each type of signals. The data for type 1, 2 and 4 originate from data set A while the third type (non-coherent waves) originate from data set C. The second row displays the data distribution from the raw data region, the displayed sigma value refers to the standard deviation of the distribution. In the third row, we show the Fourier transform for all the channels in each example region, the maximum frequency corresponds to the Nyquist limit (that is, half the sampling rate, or 250\,Hz), while the minimum frequency is defined as the inverse of the total time length of each channel's time series, which corresponds to 200 samples divided by 500\,Hz sampling rate, that is 2.5\,Hz. Finally, in the final row, we show the average spectral amplitude across channels for each frequency bin. The reference frequency at 40\,Hz, represented by the black dashed vertical line, will be used to determine a set of criteria for signal selection (see main text).}
\label{freq_content}
\end{figure}

The second data metric used to discern between noise and surface wave energy is their frequency content, which is distinct for each signal type. In Fig. \ref{freq_content}, we show the four types of signal that can be found in this DAS data set based on their distribution and spectral characteristics. The identification of these signals was made after months of careful visual inspection of the raw data. Each type can be described as follows:

\subsubsection{White noise signals}

Those mostly represent instrumental noise, which is present in all measurement channels and is characterized by a well defined narrow Gaussian distribution with a $\sigma$ broadness of less than 200. Regarding frequency content, such regions mostly contain high frequencies (above 60\,Hz), thereby explaining the linear increase towards higher frequencies of the average spectral amplitude across channels. This feature provides a strong selection criteria to identify purely white noise signals.

\subsubsection{Coherent surface waves}

Those are the target signals with useful energy that we aim to identify. Their distribution is very similar to white noise signals, which is mostly due to the rapid decrease in amplitude of the surface waves as they travel away from the source (at the apex of the "v-shape"). This behaviour makes the broad Gaussian component typical of surface waves weak, and therefore the narrow Gaussian component appears more significant. However, surface wave signals have a narrower frequency content, with a peak in average spectral amplitude at frequencies below 40\,Hz. 

\subsubsection{Non-coherent interfering waves}

If the location of the fiber's section where data are recorded is close to an area with a lot of anthropogenic activity, many sources of vibrations and interfering signals will result in a very complex pattern, difficult to interpret. Although interfering waves may not necessarily lead to non-useful signals, they can impact the overall coherency of the data and therefore make the recording less suitable for analysis. Because the distribution and frequency content of these interfering waves are very similar to those of coherent surface wave signals, it is even more challenging to distinguish them.

\subsubsection{Saturated signal}

Such signals can be seen when heavy loads (e.g. trains, trucks) are moving directly besides the section of fiber under analysis. Because of the extremely high amplitude of these signals, the recording is saturated and the resultant seismic signal is not useful. However, the high amplitude values make it relatively easy to identify (and therefore discard) such regions as they have large (above 1,000) sigma values in their distribution.

\subsection{Note on signal contamination}

Unlike most non-scientific images used for deep learning training, which can have multiple well-localized features in a single image, scientific images produced from DAS data can have mixed/merged signals making them more challenging to identify. Contamination from other signals also plays an important role in making real DAS data complex and hard to explore. For instance, any regions of any size will always be contaminated by white noise signals which is characteristic of the instrumental noise. Depending on the size of the region, some or all the signals can be present. For instance, in Fig. \ref{freq_content}, coherent surface wave signals can be found at the edges of both non-coherent interfering wave and saturated signal regions.

\section{Deep Supervised Learning}

Because of the complexity of the data, the need for an efficient method to identify useful energy signals in such a large amount of data has become urgent. Due to the nature of the DAS technology, allowing us to treat seismic data as two-dimensional images, an image classification approach using supervised machine learning becomes an evident option. Such approach consists of feeding an input 2D image into a series of regression models (or "neural network") and train the network such that the values of its outputs (or "classes"), converted into probabilities through an activation function, are tuned via a loss function to match the index value of the assigned label(s) of the input image.

\subsection{Building the Training data sets}

\subsubsection{Systematic approach}

The labeled training set was built by looking randomly at non-overlapping 200 samples x 200 channels data regions from data sets A and B, and assigning a single label (either "noise" or "waves") based on the following criteria:

\begin{itemize}
\item If the maximum average spectral amplitude below 40\,Hz is less than the minimum average spectral amplitude above 40\,Hz, the region will be assigned the "noise" label.
\item If the maximum average spectral amplitude below 40\,Hz is at least twice the mean value of all average spectral amplitudes across channels from 40 to 250\,Hz, then the region can be safely labeled as "waves".
\end{itemize}

\begin{figure}[ht]
\includegraphics[width=\columnwidth]{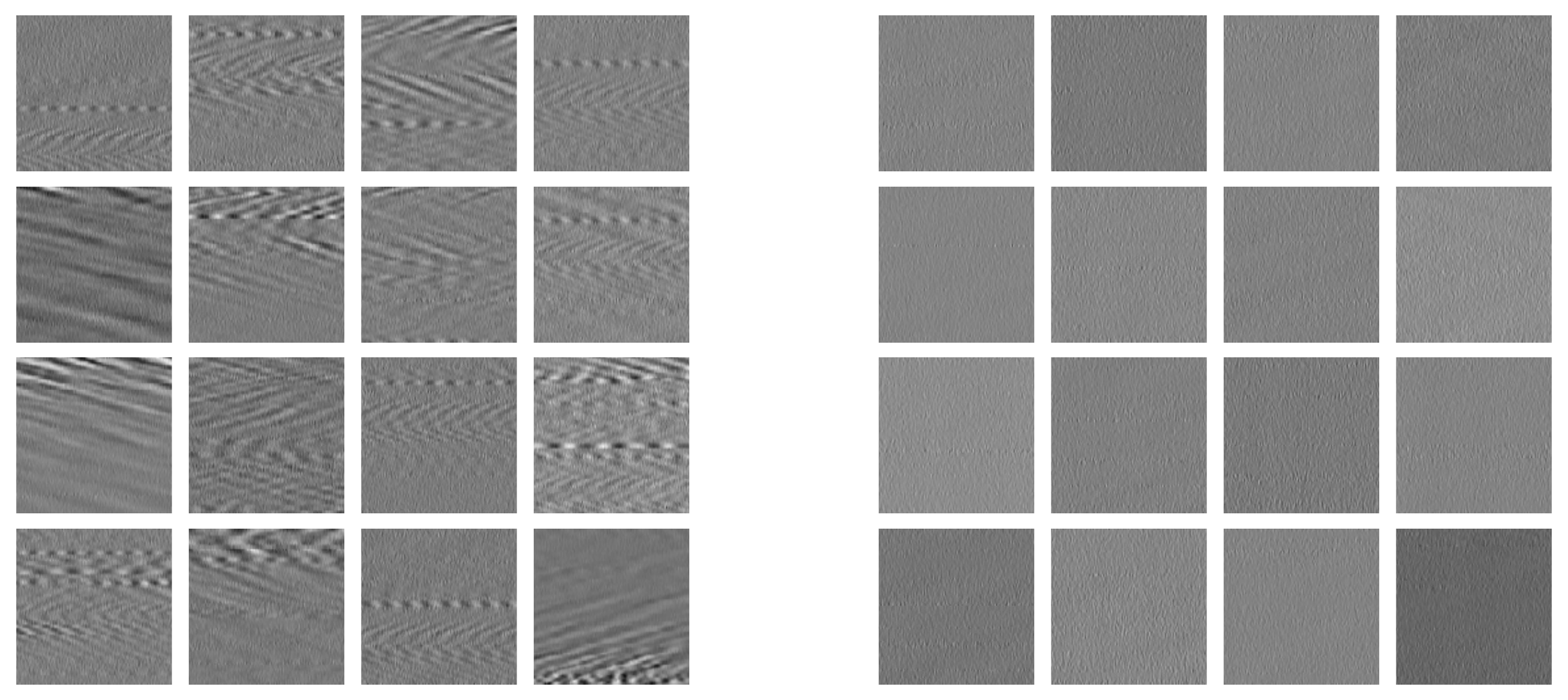}
\caption{Examples of labeled training images. The 16 images on the left are "waves" signals while the images on the right correspond to "noise" labeled images. Those regions were identified using a systematic search across all data from data sets A and B.}
\label{data sets}
\end{figure}

Regions with saturated signal, characterized by high sigma values (above 1,000), were discarded. Also, because the section of the fiber where data sets A and B originate from are located in an area where only limited activities occur at the surface, any non-saturated surface wave signals detected in the data remain coherent and can be deemed as "useful" energy signal. Those data sets allow us therefore to safely avoid complicated, non-coherent signal patterns and generate valid and un-biased training sets. In Fig. \ref{data sets}, we show some examples from both classes. The images were saved as single-channel (gray-scale) images.

A total of 174,000 training images were created for each label, and saved under a predefined folder structure readable by the PyTorch's \textsc{ImageFolder} class. The PyTorch's \textsc{subset} class will also be used to build data loaders of different sizes, allowing us to investigate the model's ability to generalize as the data set's size grows.

\subsubsection{Clustering approach}

\begin{figure}[ht]
\includegraphics[width=\columnwidth]{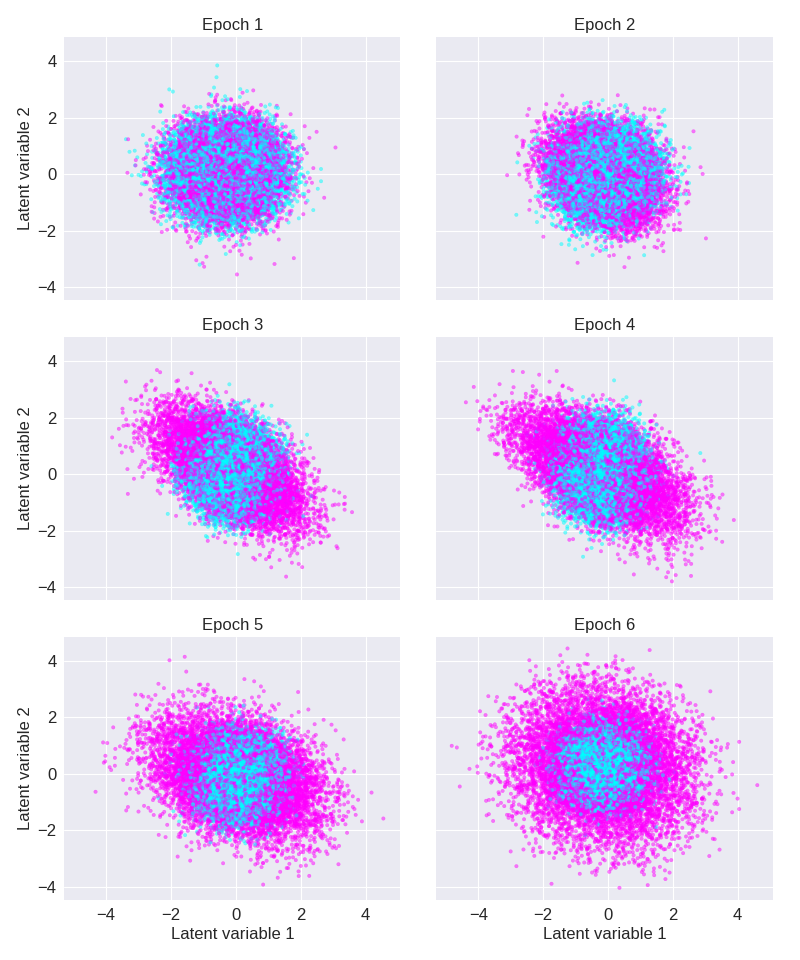}
\caption{Latent space representation from a Variational Auto-Encoder (VAE) network at different training epochs. The VAE was trained on the binary labeled data set created using the systematic search approach so one can visualize how known distinct patterns (noise versus waves) get rearranged. The blue dots correspond to "noise" labeled images. Images represented by pink dots correspond to "waves" regions. While the algorithm failed in creating 2 distinct clusters, the cluster formed by the noise images is substantially smaller than for wave images and suggest that faint wave images, clustered around the (0,0) reference point, are indistinguishable from the noise.}
\label{clustering}
\end{figure}

In an attempt to reduce human input to the minimum, a clustering algorithm using a shallow (2 hidden layers) Variational Auto-Encoder (VAE) architecture was designed with the aim of automatically creating labeled data sets by estimating the Probability Density Function (PDF) of the training data. In Fig. \ref{clustering}, we used the labeled training images to visualize the embeddings in a 2-class latent space and investigate how they rearrange over 6 epochs. This preliminary approach failed in creating separate clusters in the latent space. However, even though both clusters are superimposed, we found that white noise images get more localized than coherent surface wave signals. One possible explanation is that faint surface wave signals are still difficult to distinguish from white noise and a more complex (deeper) unsupervised learning method should be used to disentangle the differences. Finally, we not the metric learning incorporated in a VAE type of model is different than the heuristic labelling used in the systematic approach and based on physics-based statistical metrics, thereby making the learning assumptions in both cases hardly comparable.

\subsection{Neural network architecture}

Classification is done using a Residual Neural Network architecture (ResNet) \cite{resnet}. Deep convolutional neural networks are indeed efficient for image denoising and uncovering hidden structures from image datasets. The network's input is a two-dimensional (2D) single-channel tensor representing a 200x200 region of DAS data. The tensor is then fed to a series of convolutional, batch normalization and ReLU activation layers which can be divided into 4 stages. During the first stage, the input single-channel tensor is converted into a 16-channel tensor with same dimension (that is, 200x200). Each of the 16 channels represent an altered version of the original image, all being independently tuned during the training process. In a next stage, the number of channels is doubled, raising to 32 channels. Additionally, all the images are downsampled and reduced to 100x100 data regions. The same process is repeated during a third stage where the number of channels is increased to 64 and the image dimension is reduced down to 50x50. The final stage consists of an average pooling layer to reduce each of the 64 channel's 50x50 image down to one single value, and a fully connected layer to further reduce the now 64-value tensor down to the desired number of classes, that is, the total number of labels used in the training data set. Different neural network depths can also be used. In this work, we tested the algorithm with depth of 8, 14 and 20 hidden layers. The depth will essentially define the number of convolutional layers used within each stage to tune each channel's images.

\section{Results and Performance assessment}

\subsection{\textsc{mldas} software}

In order to perform this work, we have built a set of data exploration and machine learning tools specifically designed for Distributed Acoustic Sensing data\footnote{\url{https://ml4science.gitlab.io/mldas/}}. The program is available in a form of a Python package called \textsc{mldas}, publicly available through the Python package manager \textsc{pip}. The high-level structure of the software is similar to the Machine Learning benchmarking suite built by the National Energy Research Scientific Computing Center (NERSC) \footnote{\url{https://docs.nersc.gov/machinelearning/benchmarks/}} where the user can easily specify a set of training parameters in the form of a YAML script which can be called by the main training routine and executed through a SLURM script (also made available) on the NERSC Cori cluster.

\subsection{Distributed training}

The software was designed to work efficiently on large-scale parallel systems using the multi-node \textsc{DistributedDataParallel} layer in PyTorch. In order to ease the distributed training on multiple nodes, we use the \textsc{DistributedSampler} object on the data loader to ensure proper partitioning of the data set across the distributed environment, that is to ensure each copy of the model (one for each node) is trained on a different subset of data.

\begin{figure}[ht]
\includegraphics[width=\columnwidth]{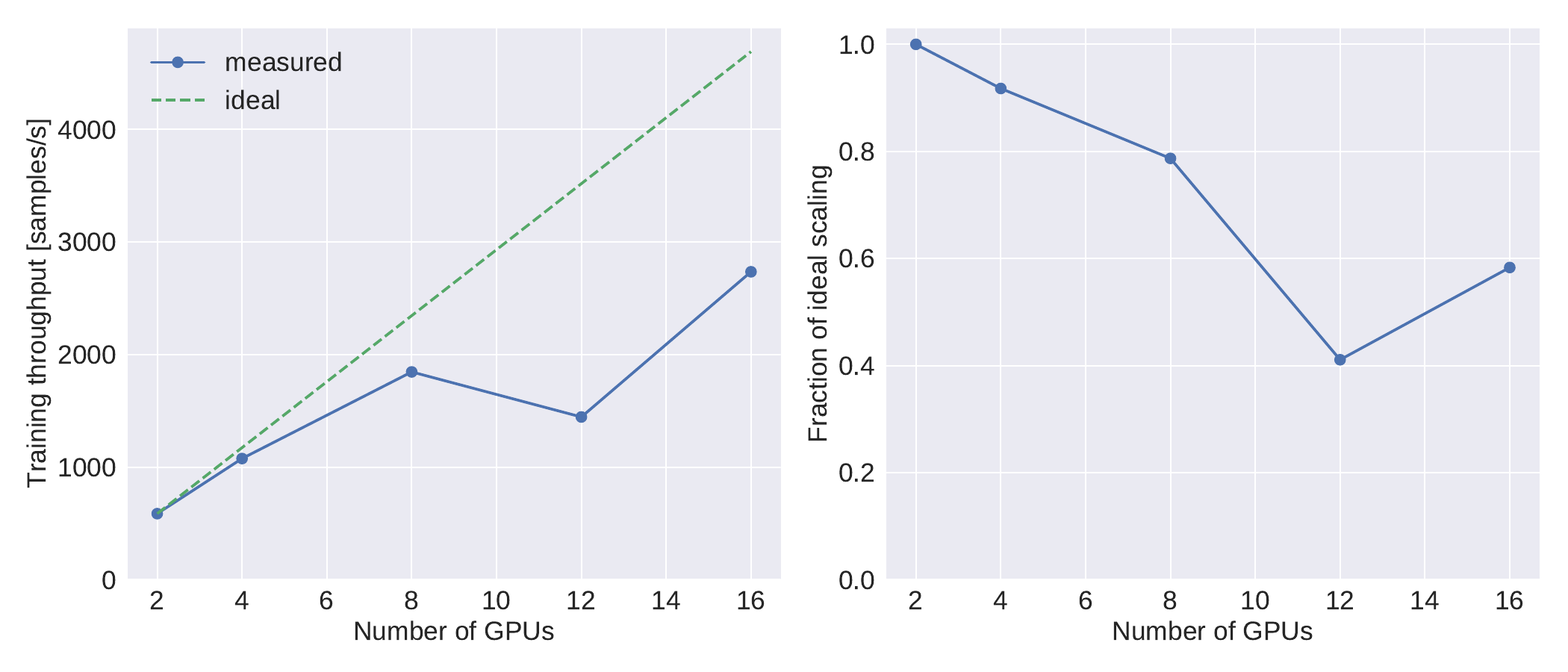}
\caption{GPU scaling results on NERSC Cori system. The left figure shows the evolution of the training throughput across GPU usage. The throughput here corresponds to the processing speed of the network, that is, the total number of samples processed per second. The right figure shows the fraction of ideal scaling, which is simply the throughput divided by the ideal scaling relative to lowest GPU rank (green curve in the left plot).}
\label{scaling}
\end{figure}

In Fig. \ref{scaling}, we show the evolution of the training throughput and fraction of ideal scaling across multiple GPUs (Graphical Processing Unit). Although the measured training throughput is below its ideal curve, a quasi-linear scaling can still be observed as we increase the total number of GPUs. There is one exception though, when using 12 GPUs the scaling does not work quite well. This suggests that if we use more than a single GPU node\footnote{Each GPU node on the NERSC Cori system contains 8 NVIDIA V100 ("Volta") GPUs}, the most efficient way to take advantage of a distributed multi-GPU system is to use all the GPUs from the additional nodes. That is, above 8 GPUs, the number of GPUs to use should be a multiple of 8.

\subsection{Hyperparameter tuning and model selection}

Different combinations of hyperparameters were explored with the aim of identifying the best model which will provide the most accurate and trustworthy probability values on unseen data. Three parameters were tuned: (1) the depth of the neural network, (2) the size of the training data set, and (3) the learning rate. Each of those parameters are assessing specific aspects of the training. The depth provides insight into the complexity of the patterns under study (a deeper network will tend to look at a deeper representation of an image). The size of the data set can ensure better generalization of the model. Finally, tuning the learning rate can help achieve faster convergence. Optimization techniques, such as adaptive batch size \cite{yao2018} or learning rate scheduler, were implemented in the early version of the training. However, because this work is more focused on building a tool that identifies useful energy in DAS data rather than making this process faster, we decided to leave the optimization steps as optional so as to demonstrate that the technique still works without the need of any optimization.

\begin{figure}[ht]
\includegraphics[width=\columnwidth]{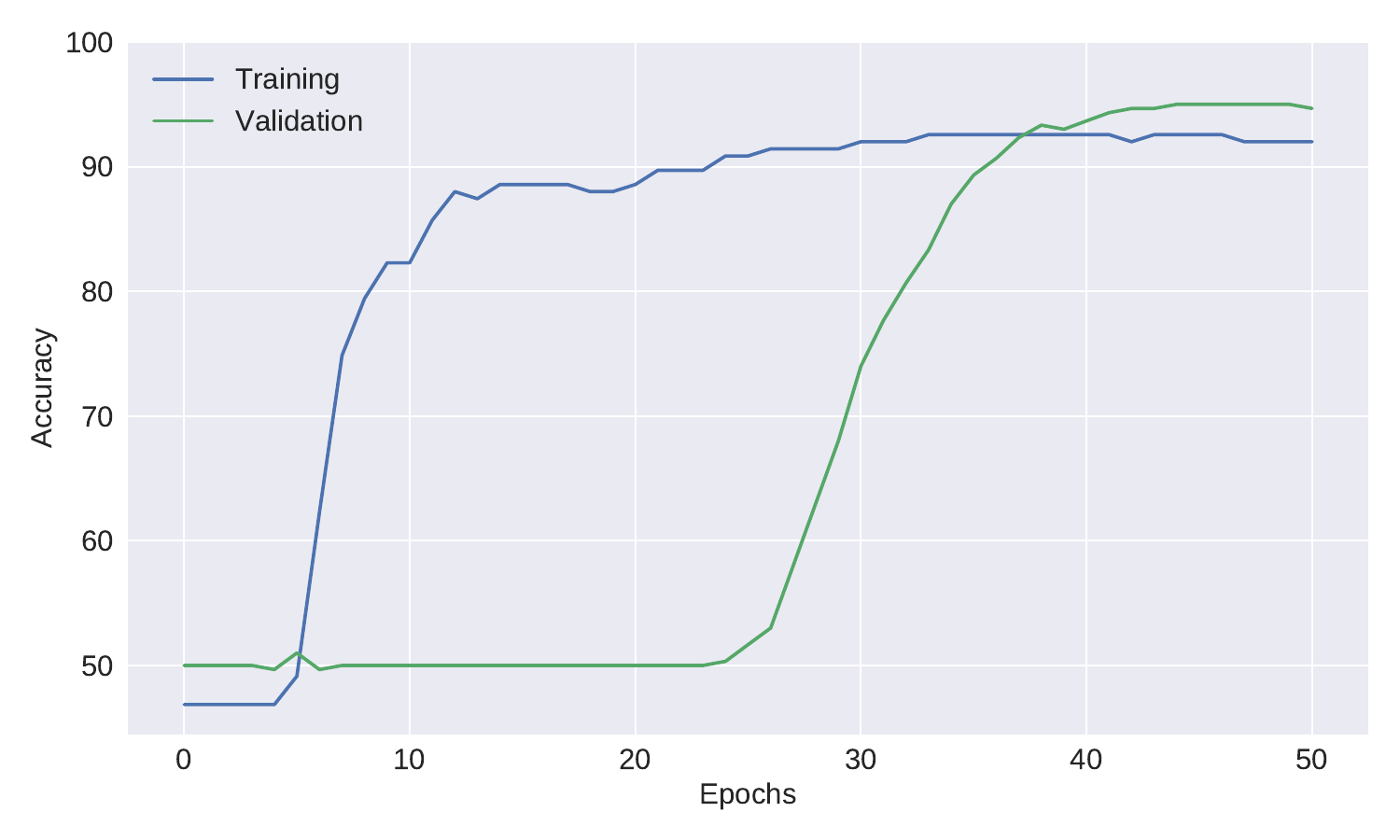}
\caption{Evolution of both training and validation accuracies for the best model over 50 epochs. The validation takes four times more epochs to reach its highest point than the training step. The higher final accuracy reached by the validation, compared to the training, can be explained by the presence of arduous cases to learn in the training data set or, conversely, easier cases to predict during validation.}
\label{best_model}
\end{figure}

The best model was found after training for 50 epochs a 14-layer deep residual neural network with a learning rate of 0.001 on a 100,000-image training data set. In Fig. \ref{best_model}, we show the training loss and validation accuracy across epochs. With a final loss of 0.19037 and final accuracy of 94.67\%, this model provides the most trustworthy results, especially for contaminated DAS regions where the model give adequate probabilities based on the presence (or not) of coherent surface wave signals.

\subsection{Inference and probability mapping}

The trained classifier can then be used to identify regions of useful coherent energy in unseen data. data set C, containing 10 days' worth of DAS data, was used to assess the quality of the training model. Each file corresponds to a 200x30000 data array (30,000 samples across 200 channels), which can be converted into 150 individual, non-overlapping, 200x200 black-and-white images. The input tensor of dimension [150,1,200,200] can then be fed to the trained model, set in evaluation mode, to output probabilities for each consecutive 200-sample region. The process on a single CPU core takes about 3 seconds for one file to complete and over half a day for the whole data set. However, using the \textsc{mpi4py} library and taking advantage of the NERSC Cori cluster, we were able to parallelize the process and reduce the inference time down to 30 minutes across 96 cores.

\begin{figure}[ht]
\includegraphics[width=\columnwidth]{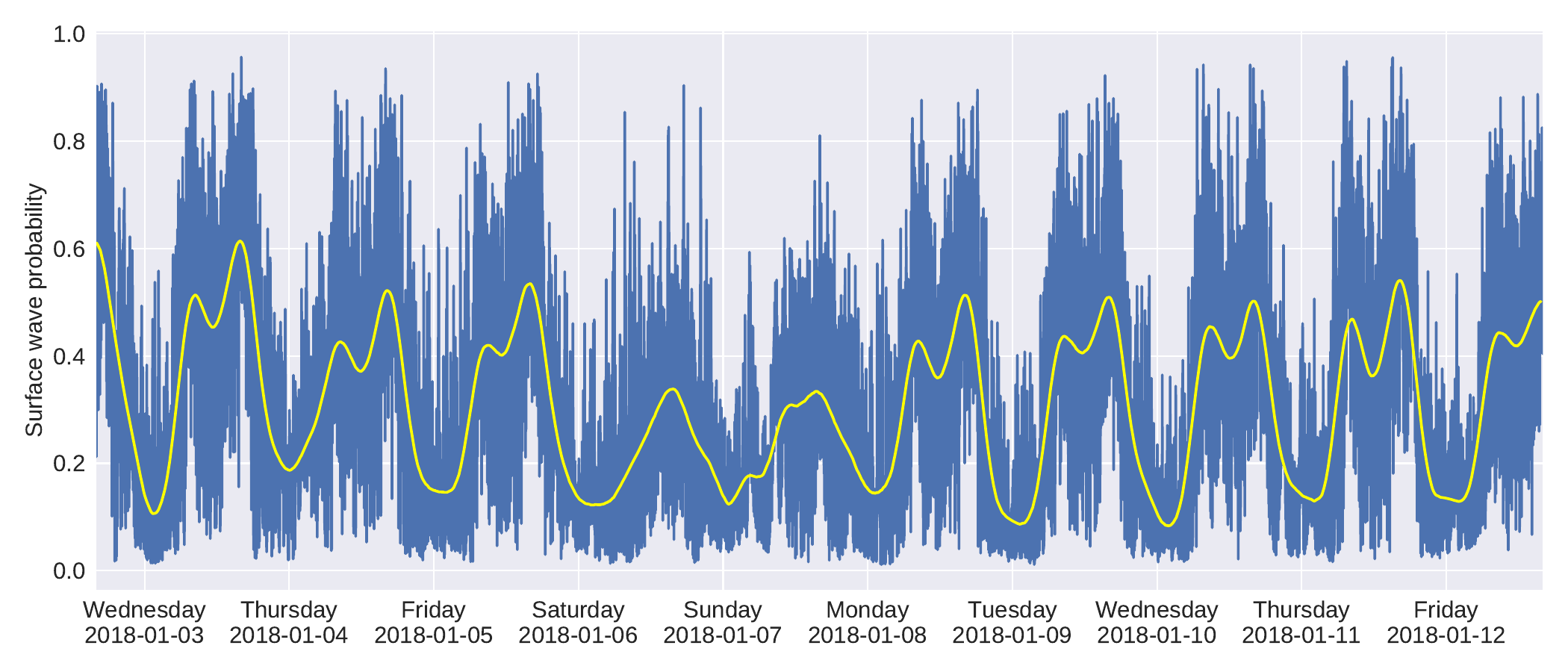}
\caption{Changes in surface wave occurrence probability over 10 days. Interpretable patterns can be found, describing the interaction of ground-based anthropogenic activities with the sensors. Data in blue show the average probability for each of the 14,400 one-minute DAS data files from data set C. The fluctuations in yellow represent the same data downsampled by a factor of 10 and convolved with a Gaussian kernel.}
\label{dailyprob}
\end{figure}

In Fig. \ref{dailyprob}, we display the change in probability of occurrence of surface waves over several days. This gives us insight into the amount of traffic occurring at the surface and hence is a way of qualitatively assess the success of our model. Low values of surface wave occurrence probability during night-time is explained by the near-absence of traffic during that period. The peaks of probability are found during day-time both in early morning and late afternoon, correlating with the time at which traffic is at his highest (that is, when people go to and come back from work). Finally, a shallow drop in surface wave probability can be seen at noon, matching with lunch time and therefore a decrease in traffic.

\begin{figure}[ht]
\includegraphics[width=\columnwidth]{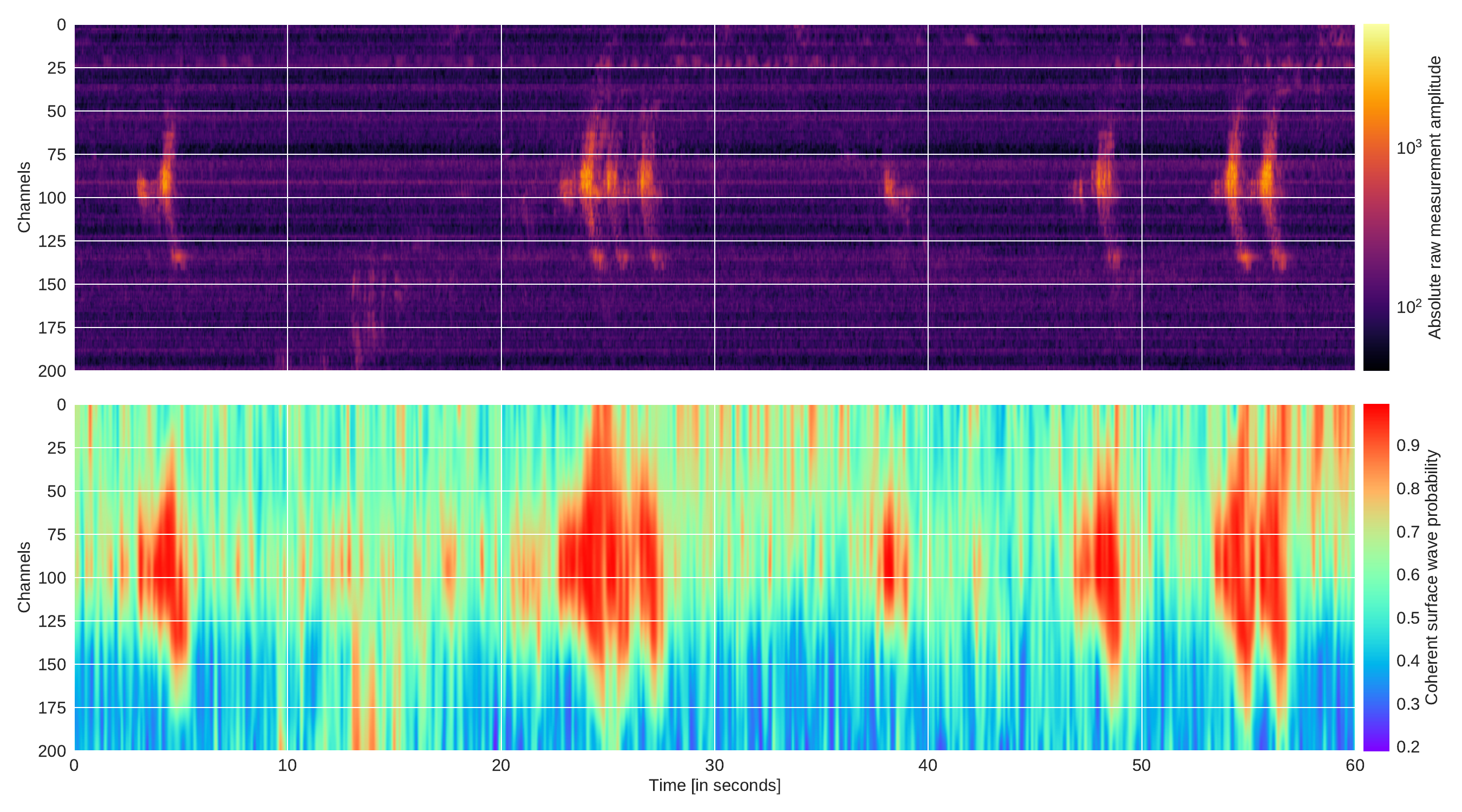}
\caption{Raw data (top) and corresponding probability map (bottom) for a 1-minute data recording. Bright signals in the top panel represent multiple car crossings where coherent surface wave signals can be found. While only highly energetic signals can be observed in the raw data, a much larger quantity of fainter but yet useful energy signals are able to get identified using our deep learning classifier. The probability map was achieved by training a prior model on smaller 50x50 regions and splitting the raw data into overlapping 50x50 regions during inference to obtain higher resolution in the channel dimension.}
\label{probmap}
\end{figure}

Figure \ref{probmap} shows an example of output probability map from a 1-minute 200-channel DAS data file. While only signals with sufficiently high energy can be observed in the raw data, the output probability map, on the other hand, does show the presence of fainter coherent wave signals detected all throughout the 1-minute region with a corresponding coherent surface wave probability higher than 0.5. This demonstrates that we are now able to efficiently identify where useful coherent energy signals are located in our big data set.

\section{Conclusion \& Future Work}

Labeled images were generated using suitable data metrics identified during an exploratory data analysis where physics insight was used to interpret the main characteristics of the DAS data. A scalable multi-GPUs deep learning technique was designed to identify regions of coherent energy in large-scale Distributed Acoustic Sensing (DAS) data, thus giving the geophysics community the ability to get the full benefit out of the large amount of data at their disposal. The convolutional residual neural network used in this work was trained on a total of 100,000 labeled images and used to infer coherent surface wave probability map on 170\,GB of DAS data under 30 minutes using 96 CPU cores.

The ability to automatically detect and quantify usable surface energy in massive, noisy DAS data sets will have important implications for the geophysical community working with this novel sensing technology. One of the most direct applications would be optimization of data selection approaches for long-term ambient noise studies. By identifying which periods of time and spatial sections contain the most usable energy, large amounts of non-useful data could be discarded and data sets could be substantially reduced. Besides, this ML-based quantification strategy could potentially be used to develop weighting schemes that could be incorporated s approaches to maximize the use of coherent energy. This technique could also be applied in earthquake-related studies, in which surface wave energy can obscure seismic events of interest. In this case, this technique would be used to discard data sections rich in infrastructure noise. 




\small
\bibliographystyle{IEEEtranN}
\bibliography{references}

\end{document}